\documentclass[pre,amsmath,superscriptaddress,showpacs,showkeys,twocolumn]{revtex4-1}
\usepackage{amssymb,amsmath}   
\usepackage[dvips]{graphicx}   
\usepackage{verbatim}   
\usepackage{color}      
\usepackage{subfigure}  
\usepackage{hyperref}   
\usepackage{gensymb}
\usepackage{epstopdf}
\usepackage{natbib}
\usepackage{enumerate}
\usepackage{mathbbol}
\usepackage{braket}
\usepackage{ulem}

\begin{document}

\title{Anomalous low-energy properties in amorphous solids and the interplay of electric and elastic interactions of tunneling two-level systems}

\author{Alexander Churkin}
\affiliation{Department of Software Engineering, Sami Shamoon College of Engineering, Beer-Sheva 8410802, Israel}
\affiliation{Department of Physics, Ben-Gurion University of the Negev, Beer Sheva 8410501, Israel}
\author{Shlomi Matityahu}
\affiliation{Department of Physics, Ben-Gurion University of the Negev, Beer Sheva 8410501, Israel}
\affiliation{Institute of Nanotechnology, Karlsruhe Institute of Technology, D-76344 Eggenstein-Leopoldshafen, Germany}
\affiliation{Department of Physics, NRCN, P.O. Box 9001, Beer-Sheva 84190, Israel}
\author{Andrii O. Maksymov}
\affiliation{Department of Chemistry, Tulane University, New Orleans, LA 70118, USA}
\author{Alexander L. Burin}
\affiliation{Department of Chemistry, Tulane University, New Orleans, LA 70118, USA}
\author{Moshe Schechter}
\affiliation{Department of Physics, Ben-Gurion University of the Negev, Beer Sheva 84105, Israel}
\date{\today}
\begin{abstract}

Tunneling two-level systems (TLSs), generic to amorphous solids, dictate the low-temperature properties of amorphous solids and dominate noise and decoherence in quantum nano-devices. The properties of the TLSs are generally described by the phenomenological standard tunneling model. Yet, significant deviations from the predictions of this model found experimentally suggest the need for a more precise model in describing TLSs.
Here we show that the temperature dependence of the sound velocity, dielectric constant, specific heat, and thermal conductivity, can be explained using an energy-dependent TLS density of states reduced at low energies due to TLS-TLS interactions.
This reduction is determined by the ratio between the strengths of the TLS-TLS interactions and the random potential, which is enhanced in systems with dominant electric dipolar interactions.

\end{abstract}

 \maketitle

\section{Introduction}

Understanding the low-temperature physics of disordered and amorphous materials has emerged as one of the most intriguing and challenging problems in condensed matter physics~\cite{YCC88,PRO02}. Below about $1\,$K, such systems exhibit physical properties that are not only qualitatively different from those of crystalline solids, but also show a remarkable degree of universality~\cite{PRO02,ZRC71,HS86,PWA87}. For instance, the specific heat and thermal conductivity are approximately linear and quadratic in temperature, respectively, while the internal friction $Q^{-1}$ is nearly temperature-independent and varies slightly between different materials.

This behavior of amorphous solids has been primarily interpreted with the model of tunneling two level systems (TLSs)~\cite{APW72,PWA72}, which will be referred to as the standard tunneling model (STM), suggesting the presence of atoms or groups of atoms that may tunnel between two nearly degenerate configurations. There were numerous suggestions targeted to describe the nature of tunneling systems and their universality, including the soft-potential model \cite{KGI83} and its further developments (see Ref.~\onlinecite{GKK89} and references therein), interaction-based models targeted to account for quantitative universality of TLSs~\cite{YCC88,Parshin94,BK96,BAL98,SM13}, glass-transition-based theory~\cite{LW01,LV18} and models based on the polaron effect~\cite{AMLD13,CY19}.
Similarly to the STM, all these theories account for the existence of TLSs at low temperatures and the resulting thermodynamic and acoustic properties of glasses.
Yet, their predictive value lies in their deviations from the STM, which has to be checked against experimental observations~\cite{SNB18}.

Marked examples of discrepancies between experimental results and theoretical predictions of the STM are the deviations from integer powers of the temperature dependence of the specific heat and thermal conductivity (see below), and the anomalous temperature dependence of the sound velocity and dielectric constant. The STM predicts a logarithmic temperature dependence, with a maximum for the sound velocity and a minimum for the dielectric constant, with a slope ratio of $1:-0.5$ between the slopes below and above the crossover temperature. Yet, experiments find different values for this ratio of slopes, typically $1:-1$~\cite{REH+95,SEH96,EH97,SEH98,TTP99}.

Whereas the original formulation of the STM neglects interactions between the TLSs, it became apparent that interactions play a significant role in phenomena such as spectral diffusion and phonon echoes~\cite{BJL77,BAL98,EC02}. TLS-TLS interactions lead to a reduction of the TLSs density of states (DOS) near zero energy~\cite{BAL98,EAL75,BSD80}. This reduction of the DOS scales with the ratio of the interaction strength to the disorder energy~\cite{CGY94,Burin95,SM13,CA15}, usually assumed to be much smaller than unity.

At the same time, there is a growing body of evidence for an energy-dependent DOS at low energies, of the form $n(E)\propto E^{\mu}$, with $0.1 < \mu < 0.3$. Even stronger energy dependence of the DOS in $a$-SiO was recently extracted from measurements of dielectric loss using superconducting lumped element resonators~\cite{SST15}. These findings are supported by earlier experiments which show indirect evidences for energy-dependent DOS: in deviations from STM predicted integer values for the temperature dependence of the specific heat, $C\propto T^{1+\alpha}$, and of the thermal conductivity, $\kappa\propto T^{2-\beta}$, with $\alpha,\beta\approx 0.1-0.3$~\cite{HS86,SRB73,LJG75};
and in the linewidth of optical transitions of ions and molecules embedded in glasses having an unusual temperature dependence $\propto T^{1.3}$~\cite{SPJ76,HJ83,THPH83}, which may arise due to dipolar interactions between the TLSs, assuming a DOS $n(E)\propto E^{\mu}$, with $\mu\approx 0.3$~\cite{SR87,HDL84,GE97}. In addition, a DOS $n(E)\propto E^{\mu}$ with $\mu\approx 0.3$ was recently assumed in Refs.~\cite{BJ14,FL15,GS18} in an effort to provide a theoretical explanation for the temperature and power dependence of $1/f$ noise in superconducting resonators at low temperatures (see, however, Ref.~\onlinecite{BAL15}). Still, it is not clear what the origin of such a marked energy
dependence of the TLS-DOS may be.

Here we calculate the single-particle TLS-DOS assuming TLS disorder energy being not much larger than the TLS-TLS interaction energy. At zero temperature we find the TLS-DOS to be significantly reduced, and well described by a power law, the power being approximately the ratio between interaction and disorder. Since the single-particle TLS-DOS involves the excitation energies of single TLSs in the environment of all other TLSs, it is temperature dependent. Indeed, at finite temperature the pseudo-gap at low energies closes gradually.

Intriguingly, we find that energy-dependent TLS-DOS accounts well not only for the anomalous power laws of the temperature dependence of the specific heat and thermal conductivity, but also for the anomalous temperature dependence of the sound velocity and dielectric constant. We discuss the energy dependence of the TLS-DOS within
the two-TLS model~\cite{SM13} and show that TLS-TLS interactions not much smaller than the random fields arise once TLS-TLS interactions are dominated by the electric dipolar interaction. Relation to experimental results is then discussed.

The paper is organized as follows: In Sec.~\ref{Sec:model} we introduce the generic model for TLSs, albeit allowing for an arbitrary ratio between the typical TLS-TLS interactions at short distances and the typical random field.
In Sec.~\ref{Sec:results} we first present (Sec.~\ref{SubSec:DOS}) the numerical results for the single-particle TLS-DOS for different ratios of interactions to random fields, and the resulting temperature dependence of the thermal conductivity and specific heat (Sec.~\ref{SubSec:heat}), and of the sound velocity and dielectric constant (Sec.~\ref{SubSec:velocity}). Relation to experiments is then discussed in Sec.~\ref{Subsec:experiments}.
In Sec.~\ref{Sec:TwoTLS} we discuss, within the two-TLS model, the possibility of the enhancement of TLS-TLS interactions as a result of the dominance of electric interactions over elastic interactions in amorphous solids. We then summarize in Sec.~\ref{Sec:Summary}.

\section{Model and TLS-DOS}
\label{Sec:model}

At low energies the system of interacting TLSs can be modelled by the effective Hamiltonian~\cite{APW72,PWA72,BJL77,BK96}
\begin{align}
\label{eq:quantumIsing}&\mathcal{H}^{}_{\mathrm{TLS}}=\sum^{}_{i}h^{}_{i}\tau^{z}_{i}+\sum^{}_{i}{\Delta}^{}_{0,i}\tau^{x}_{i}+\frac{1}{2}\sum^{}_{i\neq
j}J^{}_{ij}\tau^{z}_{i}\tau^{z}_{j},
\end{align}
where $\tau^{z}_{i}$ and $\tau^{x}_{i}$ are the Pauli matrices that represent the TLS at site $i$. The first term is the bias energy of the TLSs resulting from their interaction with static disorder. The total bias energy of TLS $i$ is therefore
$\Delta^{}_{i}\equiv h^{}_{i} + \sum^{}_{j}J^{}_{ij}\tau^{z}_{j}$, and the total energy of a TLS is given by $E=\sqrt{\Delta^2+\Delta^{2}_{0}}$. Within the STM one assumes that $J^{}_{ij}\ll h^{}_{i}$, and that $h^{}_{i}$ are homogeneously distributed, leading to the ansatz $P(\Delta,\Delta^{}_{0})=P^{}_{0}/\Delta^{}_{0}$ for the distribution of the bias energy $\Delta$ and tunneling amplitude $\Delta^{}_{0}$, and the corresponding density of states $n=P^{}_{0}L^{}_{0}$. Here $P^{}_{0}$ is a material-dependent constant and $L^{}_{0}=\ln{(\tilde{E}/\Delta^{}_{0,\mathrm{min}})}$, with $\tilde{E}$ being a large energy of the order of the disorder energy and $\Delta^{}_{0,\mathrm{min}}$ denotes the minimum tunneling amplitude of the TLSs. Generally, however, one allows an energy dependence of the TLS-DOS, i.e., $n(E)=P^{}_{0}(E)L^{}_{0}$.

The second term in the Hamiltonian~(\ref{eq:quantumIsing}) denotes TLS tunneling. Whereas this term is of utmost importance to dynamic properties, it has a small effect on the TLS-DOS, especially at energies $\gtrsim 10\,$mK relevant to most experiments. We therefore consider henceforth the random-field Ising Hamiltonian
\begin{align}
\label{eq:Ising}&\mathcal{H}=\sum^{}_{i}h^{}_{i}\tau^{z}_{i}+\frac{1}{2}\sum^{}_{i\neq
j}J^{}_{ij}\tau^{z}_{i}\tau^{z}_{j}\, ,
\end{align}
with $h^{}_{i}=h^{}_{0}c^{}_{i}$ and $J^{}_{ij}=c^{}_{ij}J^{}_{0}/(R^{3}_{ij}/R^{3}_{0}+C)$, where $c^{}_{i}$ and $c^{}_{ij}$ are normally distributed random variables with zero mean and unity variance, $R^{}_{ij}$ is the distance between TLS $i$ and TLS $j$, $R^{}_{0}$ is the typical distance between nearest TLSs, $J^{}_{0}$ denotes typical nearest neighbor TLS-TLS interaction, $C$ is a short distance cutoff, and $h^{}_{0}$ is the typical random field. Generally, the interaction term comprises both elastic and electric TLS-TLS interactions.

Below we discuss the energy-dependent TLS-DOS and its consequences
within a model of the Hamiltonian~(\ref{eq:Ising}), taking, however, $J^{}_{0}/h^{}_{0}$ to be not much smaller than unity - possible reason may be domination of electric dipolar interactions.

Since we consider the Hamiltonian~(\ref{eq:Ising}), thus neglecting $\Delta^{}_{0}$ in the calculation of the DOS, our numerically obtained TLS-DOS can be equivalently interpreted as $n(E)$ or $n(\Delta)$. We show below that the two interpretations give very similar results for the various quantities of interest to us.

\section{Results}
\label{Sec:results}

\subsection{TLS-DOS}
\label{SubSec:DOS}

We now calculate the TLS-DOS within the model presented by the Hamiltonian~(\ref{eq:Ising}) with $J^{}_{0}/h^{}_{0}=0.2$ and $0.3$. To demonstrate the power-law-like energy dependence of the low-energy TLS-DOS we perform Monte-Carlo (MC) simulations on cubic lattices of size $L^{3}$, with $L=8$ and $12$, and periodic boundary conditions are imposed. TLSs are placed randomly in the lattice with concentration $x=0.5$, and we choose $h^{}_{0}=10\,$K in accordance with its calculated value for KBr:CN \cite{CBS14}. We note that $h^{}_{0}$ is dictated by the TLS-strain interaction, and the particular choice we make here for its value is not essential for our results below.
The choice of lattice structure is for convenience, and the randomness of TLS positions in the amorphous solids is retained by the random dilution and by the randomness in $c^{}_{i}$ and $c^{}_{ij}$. We further note that the lattice constant $R_0$ denotes typical distance between adjacent TLSs, rather than interatomic spacing. We use a short distance cutoff equal to $R^{}_{0}$ (i.e. $C=1$) to account for the finite size of the TLSs, but decreasing the value of the cutoff has minimal effect on our results.

\begin{figure}[ht!]
\includegraphics[width=0.5\textwidth,height=0.27\textheight]{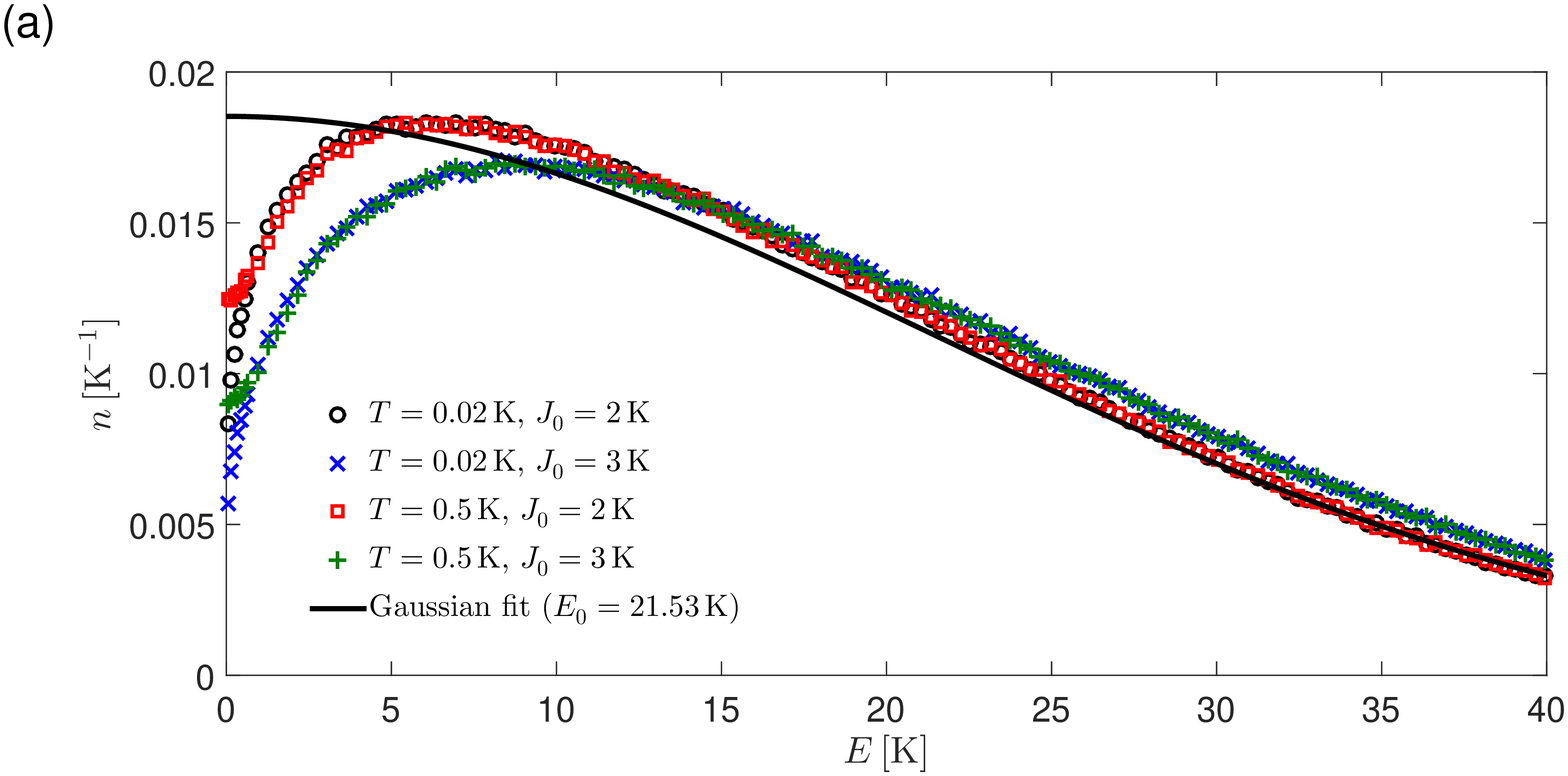}
\includegraphics[width=0.5\textwidth,height=0.27\textheight]{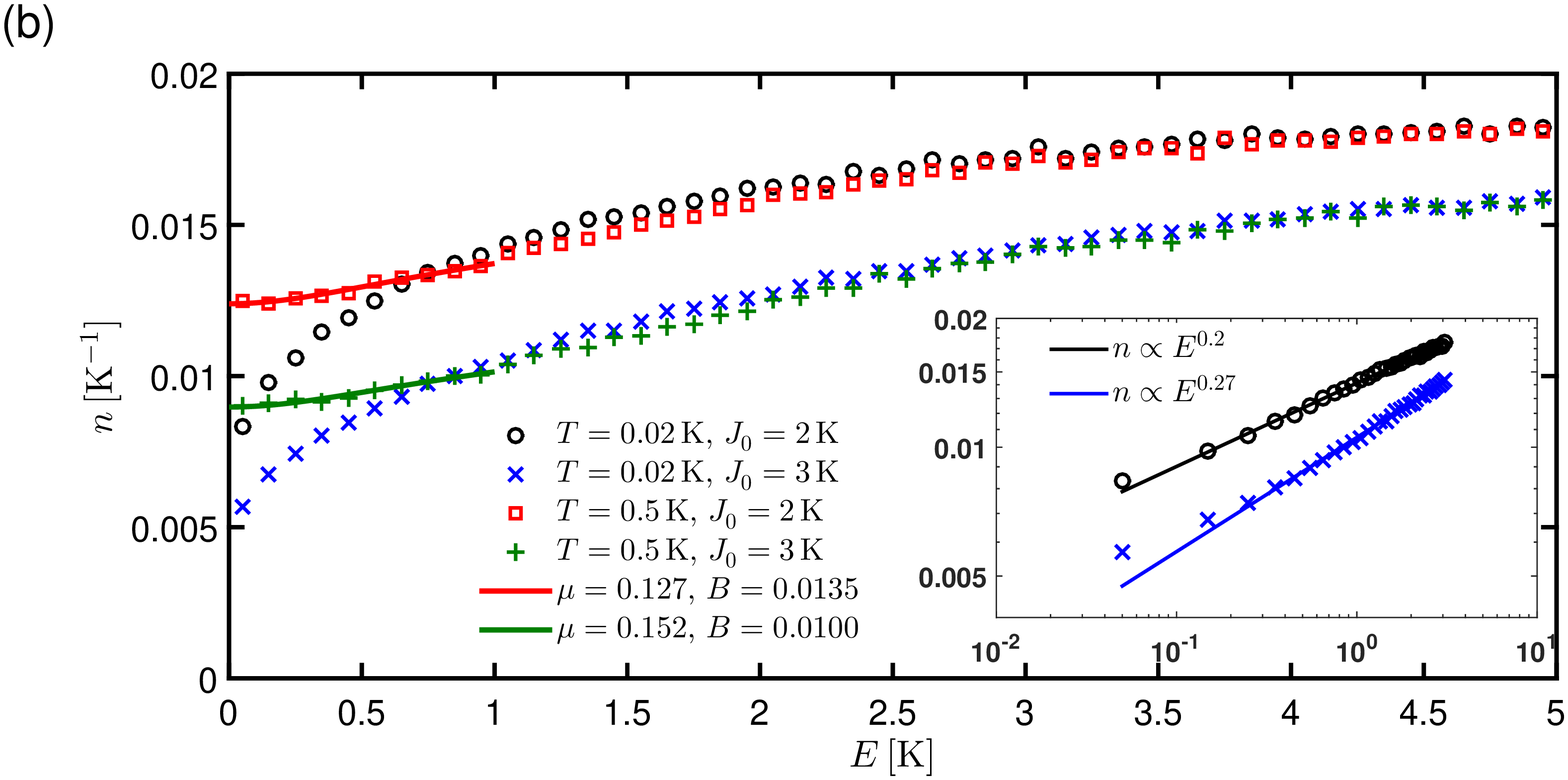}
\caption{(Color online) (a) Single-particle TLS-DOS at $T=0.02, 0.5\,$K, obtained by simulated annealing MC simulations with $L=12$ and $J^{}_{0}=2, 3\,$K ($J^{}_{0}/h^{}_{0}=0.2, 0.3$). Solid black line is a Gaussian fit (with standard deviation $E^{}_{0}=21.53\,$K) to the data at $J^{}_{0}=2\,$K, corresponding to the limit of negligible TLS-TLS interactions. (b) Zoom in to low energies. Solid lines describe low-energy fits in the range $0<E<2T$, using Eq.~(\ref{approxDOS}), for the curves corresponding to $T=0.5\,$K. Inset shows fits to the form $n\propto E^{\mu}$ for the curves corresponding to $T=0.02\,$K.
Note that the power $\mu$ decreases with increasing temperature.
}
\label{fig:1}
\end{figure}

Simulated annealing MC simulations are performed at $42$ temperatures, starting with a random realization of spins at $300\,$K and decreasing the temperature down to $0.02\,$K, running 257 MC steps at each temperature. We then reduce the temperature to $2\,\mu$K to emulate zero temperature. The single-particle DOS at a given temperature $n(T,E)$ is then calculated by measuring the excitation energies of single TLSs in a given realization, and  averaging over $10000$ independent disorder realizations for each set of parameters. While the system does not fully equilibrate within the simulated annealing technique, we verify that the final state at $2\,\mu$K is stable against single and double spin flips. This constitutes the sufficient condition for the determination of the DOS given by the Efros-Shklovskii stability criterion~\cite{EAL75,BSD80}.

In Fig.~\ref{fig:1} we plot $n(T,E)$ as a function of energy for $T=0.02\,$K and $T=0.5\,$K, interaction strengths $J^{}_{0}=2,3\,$K, and lattice size $L=12$. In the absence of interactions, the DOS is well-described by a Gaussian [solid black curve in Fig.~\ref{fig:1}(a)] with width of order $h^{}_{0}=10\,$K~\cite{SM13,CA15}. The dipolar interactions produce an Efros-Shklovskii type pseudo-gap for energies below $\sim J^{}_{0}$~\cite{EAL75,BSD80,Burin95}. As $T\rightarrow 0$, the DOS at low energies approaches a form well described by the power-law energy dependence, $n(T\rightarrow 0,E)\propto E^{\mu}$, with $\mu\approx 0.2-0.3$ [the exact value of $\mu$ depends on $J^{}_{0}$; see the inset of Fig.~\ref{fig:1}(b)]. The dipolar gap is suppressed as the temperature increases, yielding a DOS which at low energies is rather well-approximated by the function

\begin{align}
n(T,E)\approx B(T) (T^2+E^2)^{\mu(T)/2}.
\label{approxDOS}
\end{align}
The parameters $B(T)$ and $\mu(T)$ are obtained for the calculated TLS-DOS at a given temperature $T$. Fitting is performed by requiring best fitting in the energy regime $0<E<2T$. In Fig.~\ref{fig:1}(b) we plot a fit of Eq.~(\ref{approxDOS}) to the numerical DOS at $T=0.5\,$K. We find $\mu=0.1269$ for $J^{}_{0}=2$ and $\mu=0.1517$ for $J^{}_{0}=3$. We note that $\mu$ values vary little in the temperature range of $0.1-1.0\,$K, yet are smaller than the corresponding $\mu$ value at zero temperature. For a table of $B(T)$ and $\mu(T)$ values in the relevant temperature range see App.~\ref{sec:AppD}.

\begin{figure}[ht!]
\includegraphics[width=0.5\textwidth,height=0.27\textheight]{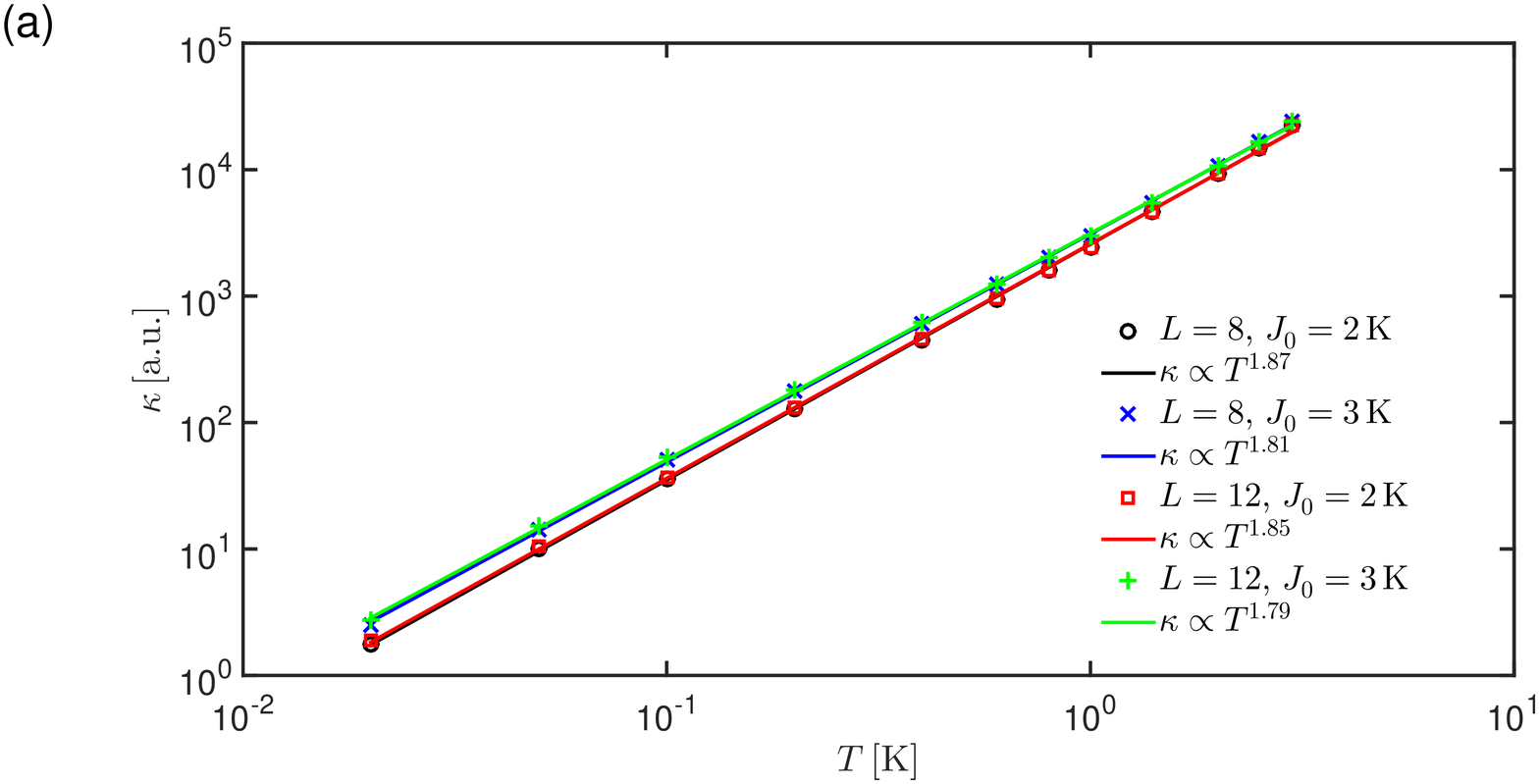}
\includegraphics[width=0.5\textwidth,height=0.27\textheight]{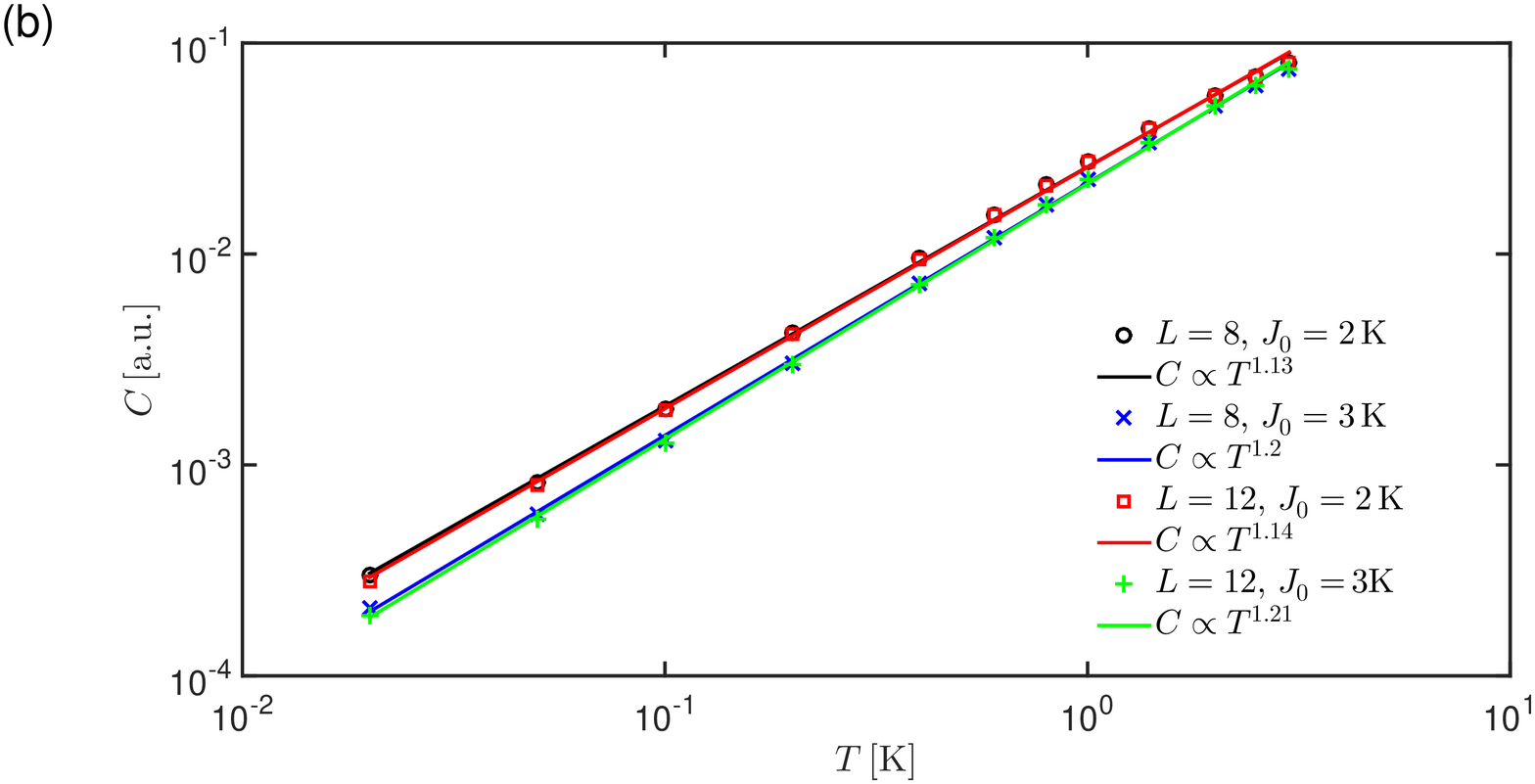}
\caption{(Color online) Temperature dependence of (a) Thermal conductivity (in arbitrary units) and (b) specific heat, obtained by Eqs.~(\ref{eq:kappa1}) and~(\ref{eq:6}) with the TLS-DOS $n(T,E)$ computed by simulated annealing MC simulations with $L=8, 12$ and $J^{}_{0}=2, 3\,$K (isolated points). Solid lines are fits to the form $\kappa\propto T^{2-\beta}$ and $C\propto T^{1+\alpha}$. Calculations correspond to a relaxed system close to equilibrium, see text.}
\label{fig:2}
\end{figure}

\begin{figure}[ht!]
\includegraphics[width=0.5\textwidth,height=0.27\textheight]{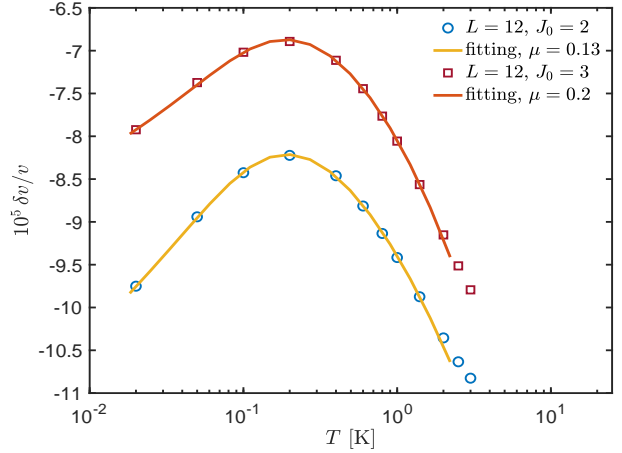}
\caption{(Color online) Temperature dependence of acoustic velocity, derived from the TLS-DOS obtained numerically from Eq.~(\ref{eq:Ising}), for 12 different temperatures, for $L=12$, $J^{}_{0}=2$ (circles) and $J^{}_{0}=3$ (squares). Solid lines are fits by the sum of Eqs.~(\ref{eq:res-gen}) and~(\ref{eq:relax-gen}), using Eq.~(\ref{approxDOS}) for the TLS-DOS with temperature independent $\mu$ as a fitting parameter. Discrepancy between the power of the calculated energy-dependent DOS [$\mu=0.2, 0.27$ for $J^{}_{0}=2, 3$, respectively, see Fig.~\ref{fig:1}(b)] and fit ($\mu=0.13, 0.2$ for $J^{}_{0}=2, 3$, respectively) is
in accordance with the reduced value of $\mu$ obtained in fitting the numerical TLS-DOS at finite temperatures using Eq.~(\ref{approxDOS}) (see Fig.~\ref{fig:1}).}
\label{fig:numericalvelocity}
\end{figure}

\begin{figure}[ht!]
\includegraphics[width=0.5\textwidth,height=0.27\textheight]{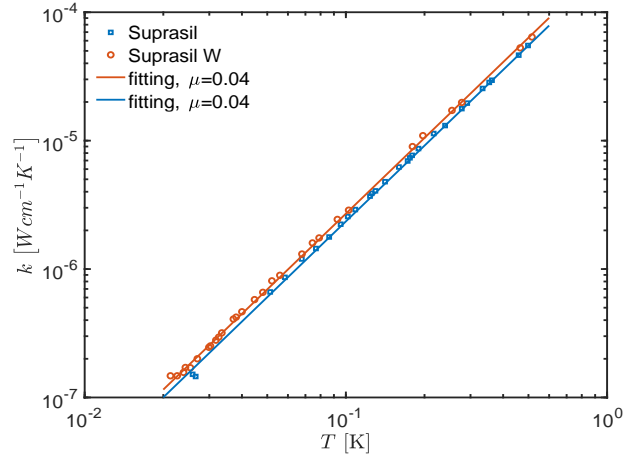}
\caption{(Color online) Temperature dependence of the thermal conductivity of vitreous silica (Suprasil and Suprasil W). Data can be well fitted to the functional form $T^{1.95}$~\cite{LJG75}. Solid lines correspond to fits by Eq.~(\ref{eq:kappa1}). The DOS is taken from Eq.~(\ref{approxDOS}) using temperature independent $\mu$ as a fitting parameter.
}
\label{fig:thermalconductivitysilica}
\end{figure}

\begin{figure}[ht!]
\includegraphics[width=0.47\textwidth,height=0.258\textheight]{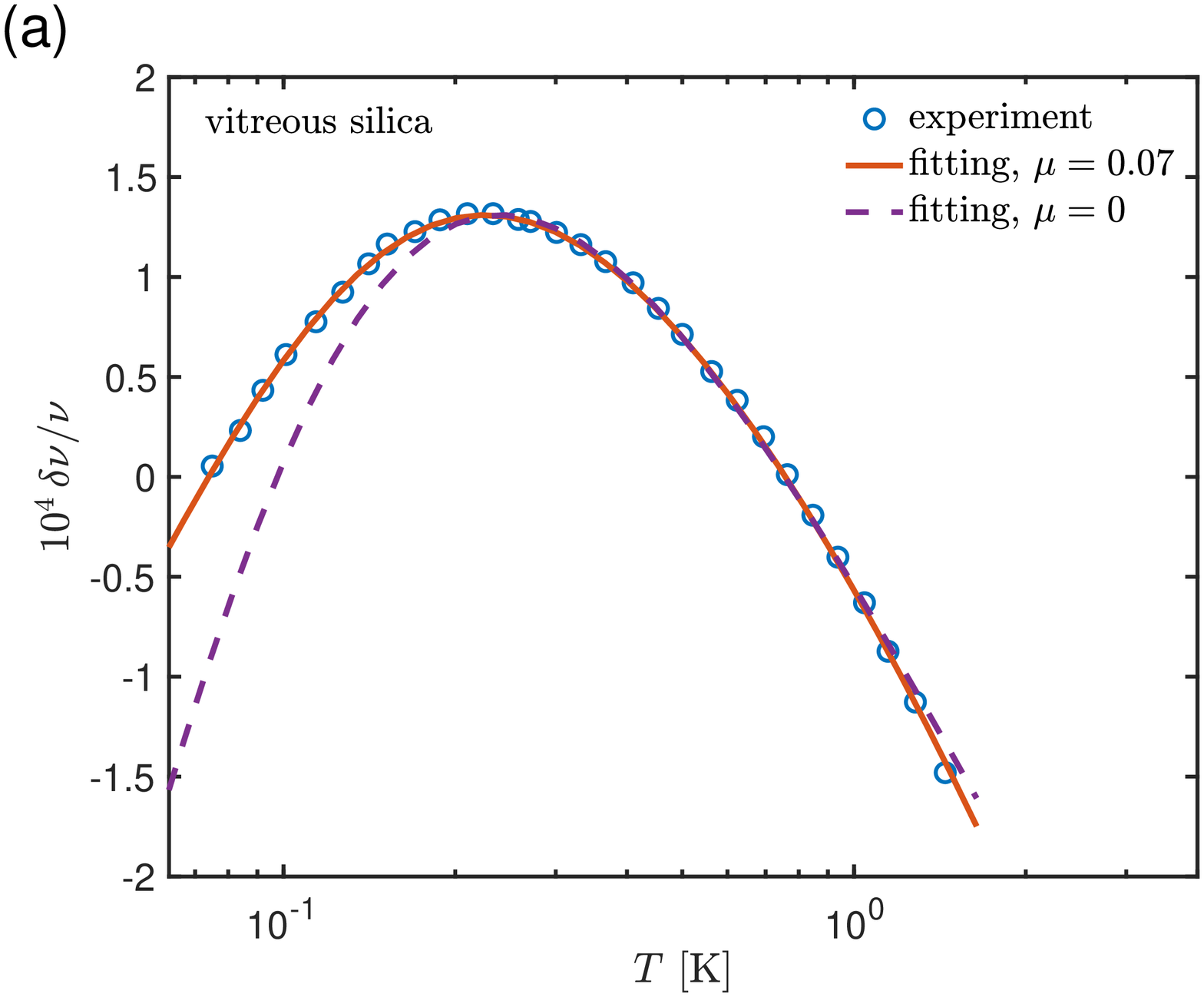}
\includegraphics[width=0.47\textwidth,height=0.258\textheight]{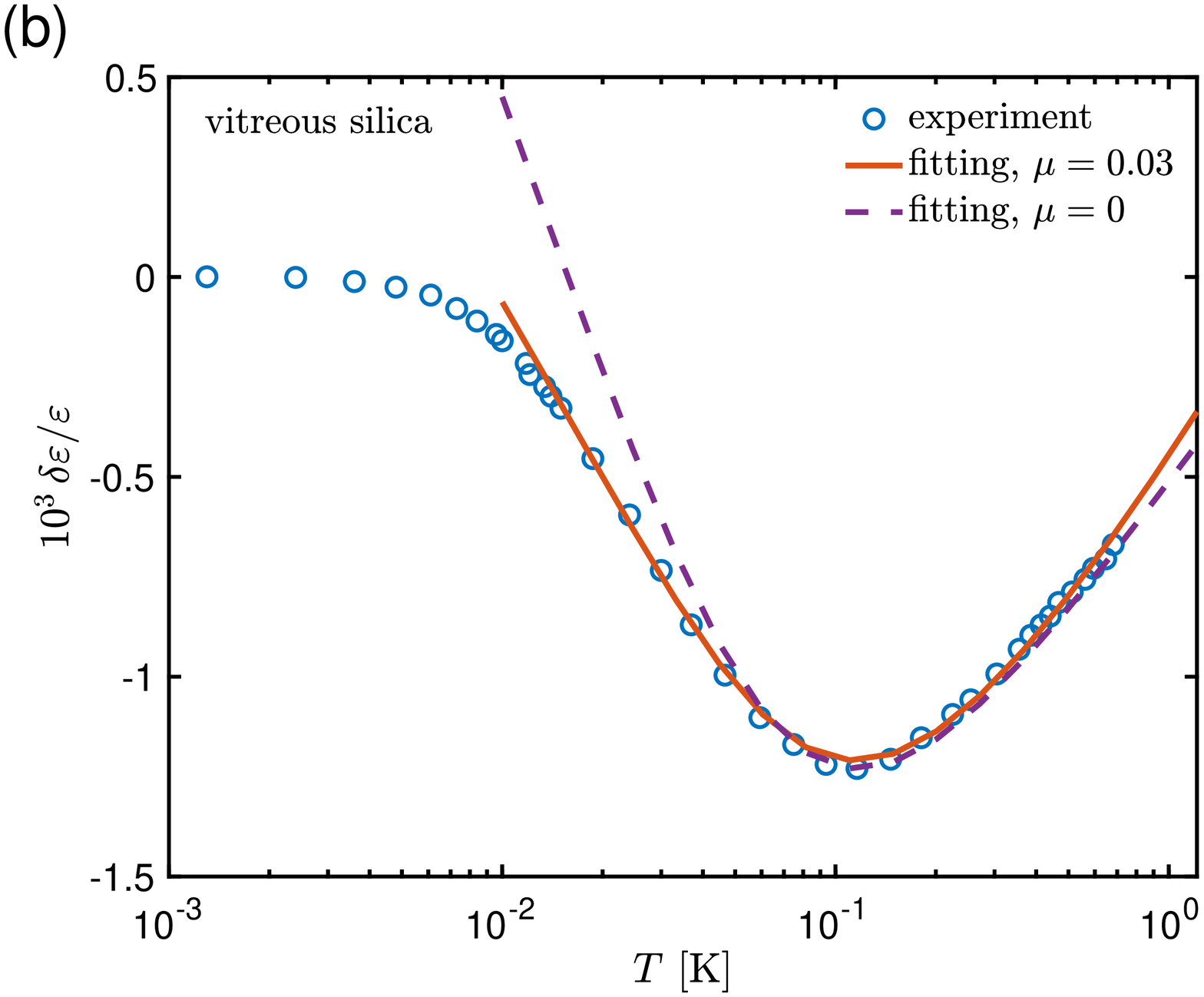}
\includegraphics[width=0.47\textwidth,height=0.258\textheight]{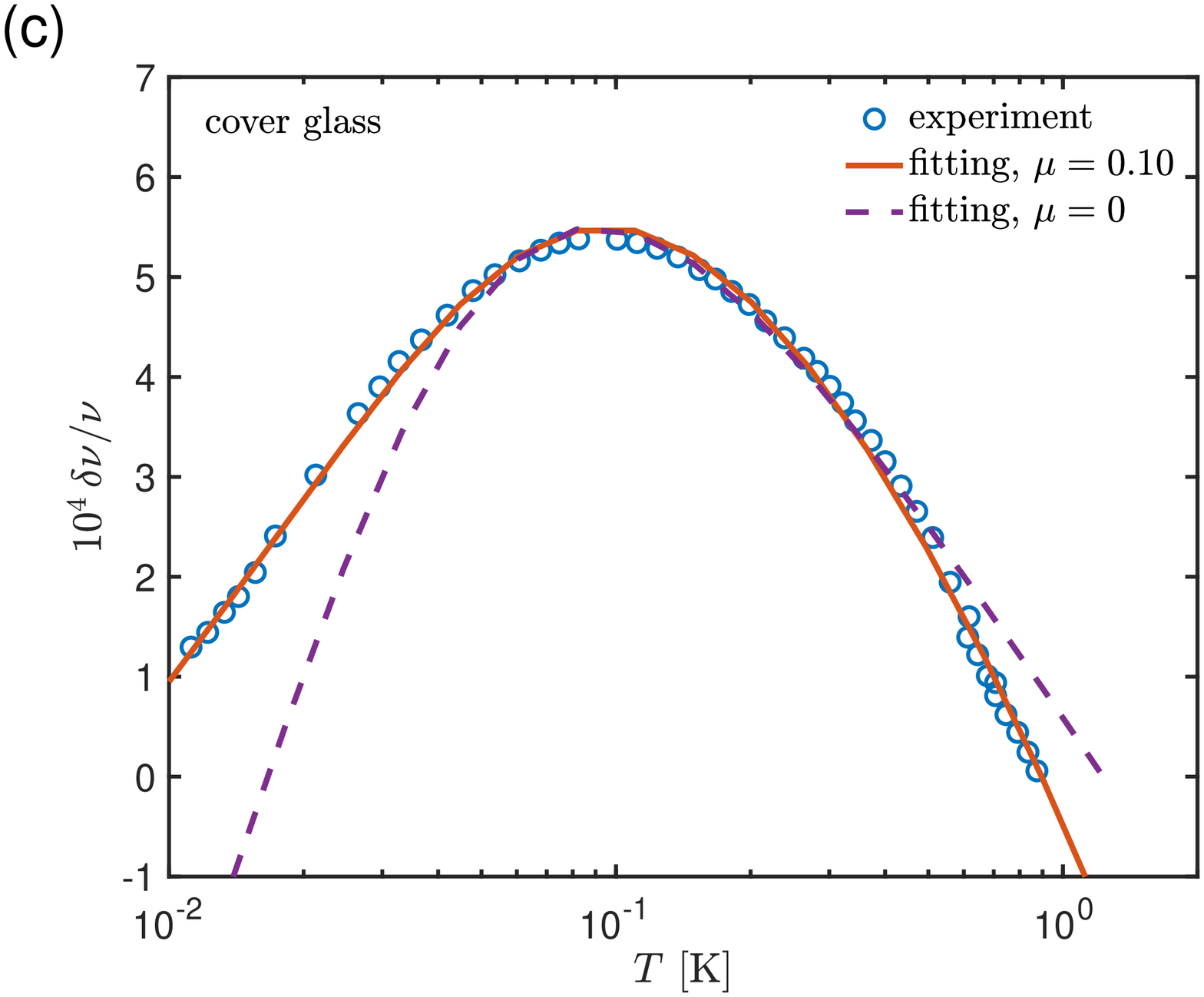}
\caption{(Color online) Temperature dependence of (a) sound velocity~\cite{TTP99} and (b) dielectric constant of vitreous silica~\cite{SEH98}, and (c) sound velocity of coverglass~\cite{CEB+94}. Solid lines correspond to fits by the sum of Eqs.~(\ref{eq:res-gen}) and~(\ref{eq:relax-gen}), using for $\omega$ the experimental values [90KHz in (a), 1KHz in (b), 5.1KHz in (c)], and using the fitting parameters $A$ [$9.4$MHz in (a), $7.9$MHz in (b), $5.6$MHz in (c)] and $P^{}_{0}\gamma^2/\rho v^2$ [$3.2 \cdot 10^{-4}$ in (a), $8 \cdot 10^{-4}$ in (b), $4.2 \cdot 10^{-4}$ in (c)]. The DOS is taken as Eq.~(\ref{approxDOS}) with temperature-independent $\mu$ as a fitting parameter. Dashed lines are best fits with $\mu=0$ as is given by the STM. }
\label{fig:approxDOS}
\end{figure}

\subsection{Thermal conductivity and specific heat}
\label{SubSec:heat}

Being well-approximated with a power-law DOS at low energies, we expect the TLS-DOS calculated from the Hamiltonian~(\ref{eq:Ising}) and plotted in Fig.~\ref{fig:1} to account well for the deviations from integer power-law exponents of the temperature dependence of the thermal conductivity and specific heat as observed in amorphous solids. Starting from the distribution function of the STM, $P(\Delta,\Delta^{}_{0})=P^{}_{0}(\Delta)/\Delta^{}_{0}$, the thermal conductivity $\kappa(T)$ is found by calculating~\cite{PWA87} $\kappa(T)=\frac{1}{3}\sum^{}_{\alpha}\int^{\infty}_{0}C^{}_{\text{ph},\alpha}(E)v^{}_{\alpha}\ell^{}_{\text{ph},\alpha}(E)dE$, where $C^{}_{\text{ph},\alpha}(E)=E^{4}/\left(8\pi^{2}\hbar^{2}v^{3}_{\alpha}T^{2}\sinh^{2}(E/2T)\right)$
is the contribution factor of phonons with energy $E$ and polarization $\alpha$ to the Debye heat capacity, $v^{}_{\alpha}$ is the sound velocity and
\begin{align}
\label{eq:5}\ell^{-1}_{\text{ph},\alpha}(E)=&\frac{\pi\gamma^{2}_{\alpha}}{\rho v^{3}_{\alpha}}\int d\Delta P^{}_{0}(\Delta)\int^{\sqrt{E^{2}-\Delta^{2}}}_{0}d\Delta^{}_{0}\frac{\Delta^{}_{0}}{E}\times\nonumber\\&\tanh(E/2T)\delta\left(E-\sqrt{\Delta^{2}+\Delta^{2}_{0}}\right)
\end{align}
is the phonon inverse mean free path due to interaction with resonant TLSs (i.e., TLSs with energy splitting equal to the phonon energy), characterized by the coupling strength $\gamma^{}_{\alpha}$, where $\rho$ is the mass density.

In order to relate $P^{}_{0}(\Delta)$ to the numerical TLS-DOS, calculated for the Hamiltonian~(\ref{eq:Ising}), we take $E=\Delta$ [i.e.\ we neglect $\Delta^{}_{0}$ in the delta function in Eq.~(\ref{eq:5})]. The expression for the thermal conductivity then reads
\begin{align}
\label{eq:kappa1}\kappa(T)\propto\int^{\infty}_{0}\frac{E^{3}dE}{T^{2}\sinh^{2}(E/2T)\tanh(E/2T)P^{}_{0}(E)}.
\end{align}
The prefactor in Eq.~(\ref{eq:kappa1}) contains material-dependent constants which are independent of temperature. To study the temperature dependence of the thermal conductivity we calculate the last integral in Eq.~(\ref{eq:kappa1}) with $P^{}_{0}(E)$ replaced by the numerically calculated TLS-DOS $n(T,E)$, and represent the thermal conductivity in arbitrary units. For a similar calculation using the interpretation of the calculated TLS-DOS as $n(\Delta)$, without performing the approximation $E=\Delta$ in Eq.~(\ref{eq:5}), see App.~\ref{sec:AppC}. The difference between the two approximations is exemplified also in the calculation of the velocity, by comparing Eq.~(\ref{eq:relax-gen}) and Eq.~(\ref{eq:relax-genDelta}); yet the results obtained differ only negligibly, as both approximations differ by logarithmic corrections.

Similarly, the specific heat $C(T)$ is evaluated, taking the Boltzmann constant $k^{}_{B}=1$, as~\cite{PWA87}
\begin{align}
\label{eq:6}&C(T)=\int^{\infty}_{0}\frac{n(T,E)E^{2}dE}{4T^{2}\cosh^{2}\left(E/2T\right)}.
\end{align}

Figure~\ref{fig:2} shows log-log plots of the thermal conductivity and the specific heat as a function of temperature, for $J^{}_{0}=2, 3\,$K and $L=8, 12$. In all cases, the thermal conductivity and the specific heat obey a power-law dependence,
$\kappa\propto T^{2-\beta}$ and $C\propto T^{1+\alpha}$, with $\alpha$ and $\beta$ in
the range $0.1-0.2$. Note that we do not consider here the slow logarithmic time dependence of the specific heat, resulting from the large variance in TLS relaxation times, that can enhance the temperature dependence of the specific heat~\cite{Meissner81,Enss05}. Our results correspond to a given long time, as the system is out of equilibrium. Taking into account the time dependence of the specific heat would therefore result in a stronger temperature dependence compared to our results here.

\subsection{Sound velocity and dielectric constant}
\label{SubSec:velocity}

Given the above mentioned long-standing discrepancy between STM predictions and experimental results, it is of interest to study the consequences of energy-dependent TLS-DOS on the temperature dependence of the sound velocity and dielectric response at low temperatures. The temperature dependence of these quantities has two contributions coming from the resonant and relaxation processes~\cite{PWA87}. Considering the sound velocity, the contribution of the resonant process is of the form
\begin{align}
\frac{\delta v^{}_{\mathrm{res}}}{v}=
-\frac{1}{L^{}_{0}}\frac{\gamma^2}{\rho v^2}\int_{0}^{\infty} \frac{n(T,E)dE}{E}\tanh\left(\frac{E}{2T}\right) \, ,
\label{eq:res-gen}
\end{align}
where $v$ and $\gamma$ are characteristic values for the velocity and for the interaction constant. For the relaxation process one has
\begin{align}
\frac{\delta v^{}_{\mathrm{rel}}}{v}
=&-\frac{1}{L^{}_{0}}\frac{\gamma^2}{\rho v^2}\int_{0}^{\infty} \frac{n(T,E)dE}{2T\cosh^2\left(E/2T\right)}
\nonumber\\
&\times\int_{0}^{1} \frac{\sqrt{1-x^2}dx}{x}\frac{1}{1+\frac{\omega^2}{\left[Ax^2E^3\coth\left(E/2T\right)\right]^2}},
\label{eq:relax-gen}
\end{align}
such that $\delta v/v=(\delta v^{}_{\mathrm{res}}+\delta v^{}_{\mathrm{rel}})/v$.
Here $\omega$ is the probing frequency and $A\equiv \omega/T^{3}_{0}$, where $T^{}_{0}$ is a crossover temperature of the order of the temperature at which the sound velocity obtains a maximum value \cite{PWA87}. The corresponding expressions for the dielectric constant $\epsilon$ are obtained by substituting $\gamma^{2}/(\rho v^{2})\rightarrow p^{2}/(4\pi\epsilon\epsilon^{}_{0})$, where $p$ is the TLS dipole moment and $\epsilon^{}_{0}$ is the vacuum permittivity (see App.~\ref{sec:AppC} for a similar calculation using the interpretation of the DOS as $n(\Delta)$).

In Fig.~\ref{fig:numericalvelocity} we plot the sound velocity calculated as the sum of Eqs.~(\ref{eq:res-gen}) and~(\ref{eq:relax-gen}), using the numerically calculated DOS for the Hamiltonian~(\ref{eq:Ising}) with ratios $J^{}_{0}/h^{}_{0} = 0.2, 0.3$, for $12$ temperatures below and above the temperature corresponding to the maximum in sound velocity. We find the ratio between the logarithmic slopes below and above the crossover temperature to be roughly $1:-1$ for $J^{}_{0}/h^{}_{0}=0.2$, and even a steeper descent beyond the crossover temperature for $J^{}_{0}/h^{}_{0}=0.3$.

\subsection{Relation to experiments}
\label{Subsec:experiments}

Our numerical results motivate us to use energy-dependent TLS-DOS to fit the experimental data for the temperature dependence of the thermal conductivity, sound velocity, and dielectric constant. Yet, such a task requires a numerical calculation of the TLS-DOS $n(T,E)$ at many values of the ratio $J^{}_{0}/h^{}_{0}$, which is rather complicated. We therefore take a simpler approach and consider the dependence of the TLS-DOS on energy and temperature as given in Eq.~(\ref{approxDOS}), allowing the power $\mu$ to serve as a free fitting parameter, albeit independent of temperature. We first test this approach by fitting our numerical results in Fig.~\ref{fig:numericalvelocity}.
As can be seen, a good fit is obtained by using Eq.~(\ref{approxDOS}), with a fixed power $\mu$, for the TLS-DOS in the sum of Eqs.~(\ref{eq:res-gen}) and~(\ref{eq:relax-gen}). We note, however, that the temperature-independent values obtained ($\mu=0.13, 0.2$ for $J^{}_{0}/h^{}_{0}=0.2, 0.3$, respectively) are smaller than the powers $\mu$ describing the numerically simulated TLS-DOS at zero temperature [$\mu=0.2, 0.27$ for $J^{}_{0}/h^{}_{0}=0.2, 0.3$, respectively; see the inset of Fig.~\ref{fig:1}(b)].

We now consider experimental data, starting with the anomalous power of the power-law functional form of the temperature dependence of the thermal conductivity of amorphous solids. An example for vitreous silica is shown in Fig.~\ref{fig:thermalconductivitysilica}. We find an excellent fit of the data by Eq.~(\ref{eq:kappa1}), using the DOS $n(T,E)$ given by Eq.~(\ref{approxDOS}), within the approximation of temperature-independent power $\mu$. Here, too, the value of $\mu=0.04$ of the fit function is somewhat smaller than the power $0.05$ one would obtain using zero temperature TLS-DOS $n(E)\propto E^{\mu}$.

We consider next the temperature dependence of the sound velocity and dielectric response in amorphous solids. In Fig.~\ref{fig:approxDOS} we show data for vitreous silica and for coverglass,
displaying the usually found ratio of approximately $1:-1$ between the slopes of the logarithmic temperature dependence below and above the crossover temperature.
Good fits are obtained by the sum of Eqs.~(\ref{eq:res-gen}) and~(\ref{eq:relax-gen}), using the DOS $n(T,E)$ of Eq.~(\ref{approxDOS}) within the approximation of temperature-independent power $\mu$.
Given the similar fitting of the numerical data in Fig.~\ref{fig:numericalvelocity}, it is suggestive that the TLS-DOS of vitreous silica and coverglass are energy-dependent, and can be described by a power-law, with a power somewhat larger than the $\mu$ values found in the corresponding fits. We further note the larger $\mu$ value for coverglass in comparison to those found for vitreous silica, which suggests a larger ratio of interactions to disorder in the former. We argue below that this difference may be related to the relative largeness of the strength of the electric dipole-dipole interactions in coverglass in comparison to vitreous silica.

An intriguing issue is exemplified in the data of Classen {\it et. al.} \cite{CEB+94} for the temperature dependence of the sound velocity taken at driving voltages of $0.7\,$V and $10\,$V. Classen {\it et. al.} argue that consistently, data taken at non-equilibrium ($10\,$V plot here) displays agreement with the $1:-0.5$ slope ratios for the logarithmic temperature dependence of the sound velocity below and above the crossover temperature, as predicted by the STM, while data taken in equilibrium shows slope ratios of roughly $1:-1$. An explanation of this observation is beyond the scope of this paper. Yet, we would like to note that within our approach, the physics behind the anomalous temperature dependence of the sound velocity, i.e., the energy dependence of the TLS-DOS, requires the system to be close to equilibrium.

\begin{center}
\begin{table*}
\begin{tabular}{| c | c | c | c |}
\hline
& \textrm{\begin{tabular}{@{}c@{}}Interaction between \\NN general defects \end{tabular}}
 & \textrm{TLS disorder energy} & \textrm{Interaction between NN TLSs} \\
\hline \textrm{Two-TLS model} & $\sim \gamma^{2}_{\mathrm{s}}/(\rho
v^{2}R^{3}_{0})\sim T^{}_{\mathrm{g}}\approx 300-1000\,$K &
$h^{}_0\sim \gamma^{}_{\mathrm{s}}\gamma^{}_{\mathrm{w}}/(\rho
v^{2}R^{3}_{0})\approx 10\,$K &
$J^{}_{0}\sim \gamma^{2}_{\mathrm{w}}/(\rho
v^{2}R^{3}_{0})\approx 0.1-0.3\,$K \\ \hline
\textrm{
\begin{tabular}{@{}c@{}}Two-TLS model with strong \\electric dipolar interactions \end{tabular}} & $\sim \gamma^{2}_{\mathrm{s}}/(\rho
v^{2}R^{3}_{0})\sim T^{}_{\mathrm{g}}\approx 300-1000\,$K &
$h^{}_{0}\sim \gamma^{}_{\mathrm{s}}\gamma^{}_{\mathrm{w}}/(\rho
v^{2}R^{3}_{0})\approx 10\,$K &
$J^{}_{0}\sim p^2/(4\pi\epsilon\epsilon^{}_{0}
R^{3}_{0})\approx 2-3\,$K \\ \hline
\end{tabular}
\caption{\label{tbl} Comparison between: typical energy scales of the interactions between nearest neighbor (NN) defects, the resulting disorder energies for the abundant ($\tau$)-TLSs at low energies, dominating low-temperature physics,
and the interactions between ($\tau$)-TLSs.
(i) (Top row) As derived by the two-TLS model with dominant elastic interactions~\cite{SM13}, (ii) (Bottom row) As presented here for the two-TLS model with strong electric dipolar interactions. Note the small TLS disorder energies in comparison to the value of $300-1000\,$K assumed by the STM.}
\end{table*}
\end{center}

\section{Electric dipolar TLS-TLS interactions}
\label{Sec:TwoTLS}

The consideration of TLS-strain interaction as an addition to the STM was introduced by Jackle~\cite{Jackle72} to explain internal friction experiments. TLS-strain interactions result in effective TLS-TLS interactions, which lead to spectral diffusion~\cite{BJL77}, and to the dipolar gap in the single-particle TLS-DOS at low energies, which at $T=0$ takes the form~\cite{BJL77,Burin95,BK96}

\begin{align}
n(E)=\frac{n^{}_{0}}{1+c\tilde{J}^{}_{0}n^{}_{0} \log{(\tilde{J}^{}_{0}/R^{3}_{0}E)}},
\label{eq:dipolargap}
\end{align}
where $c=2\pi/3$, $\tilde{J}^{}_{0} \equiv J^{}_{0}R^{3}_{0}$ is the interaction constant, and $n^{}_{0}\equiv n(E=J^{}_{0}) = 1/(h^{}_{0}R^{3}_{0})$. Phonon attenuation data at low temperatures dictates, assuming solely strain mediated TLS-TLS interactions, a value of
$J^{}_{0}/h^{}_{0}=\tilde{J}^{}_{0}n^{}_{0}\approx 0.03$~\cite{APW72,PWA72,Jackle72}, which corresponds to a near power-law energy-dependent TLS-DOS with $\mu \approx 0.03$.
In this paper we suggest that larger values of the ratio $J^{}_{0}/h^{}_{0}$ result in enhanced energy dependence of the single-particle TLS-DOS, which in turn may explain the anomalous behavior of the acoustic velocity and dielectric constant, as well as the anomalous powers of the temperature dependence of the specific heat and thermal conductivity.
We now discuss what may be a cause for an enlarged ratio of interaction strength to random field strength, and specifically the consequences of TLSs having a larger electric dipolar interaction compared to their phonon-mediated interaction.
The emergence of a larger ratio of $J^{}_{0}/h^{}_{0}$ in the presence of dominant electric dipolar interactions is naturally obtained within the theoretical framework of the two-TLS model~\cite{SM13}. We thus begin with a presentation of the main features of the two-TLS model relevant to our discussion. A more detailed discussion of the model is deferred to Appendix~\ref{sec:AppA}.

First considering only elastic interactions, the two-TLS model divides TLSs into two groups, with bimodal distribution of their interaction strengths with the strain, denoted by $\gamma^{}_{\mathrm{w}}$ for the weakly interacting $\tau$-TLSs, which correspond to the abundant TLSs at low energies, and by $\gamma^{}_{\mathrm{s}}$ for the other defects, where $g\equiv\gamma^{}_{\mathrm{w}}/\gamma^{}_{\mathrm{s}}\approx 0.02$~\cite{SM13,SM08,GS11,CBS13,CBS14,GGS15}. The Hamiltonian~\ref{eq:Ising} is then derived as the low-energy effective Hamiltonian of the system (see Ref.~\cite{SM13} and also App.~\ref{sec:AppA}), with
\begin{align}
\label{eq:Random_field_and_interactions}
&h^{}_{i}\approx\frac{c^{}_{i}\gamma^{}_{\mathrm{w}}\gamma_{\mathrm{s}}}{\rho v^{2}R^{3}_{0}} &;&
& J^{}_{ij}\approx\frac{c^{}_{ij}\gamma^{2}_{\mathrm{w}}}{\rho v^{2}
R^{3}_{ij}}.
\end{align}
Here, $R^{}_{0}$ is the typical distance between nearest two-level defects, $R^{}_{ij}$ denotes the distance between $\tau$-TLS $i$ and $\tau$-TLS $j$, and the parameters $c^{}_{i}$, $c^{}_{ij}\sim O(1)$
can be regarded as normally distributed random variables~\cite{SM08}.

\begin{figure}[ht!]
\includegraphics[width=0.5\textwidth,height=0.27\textheight]{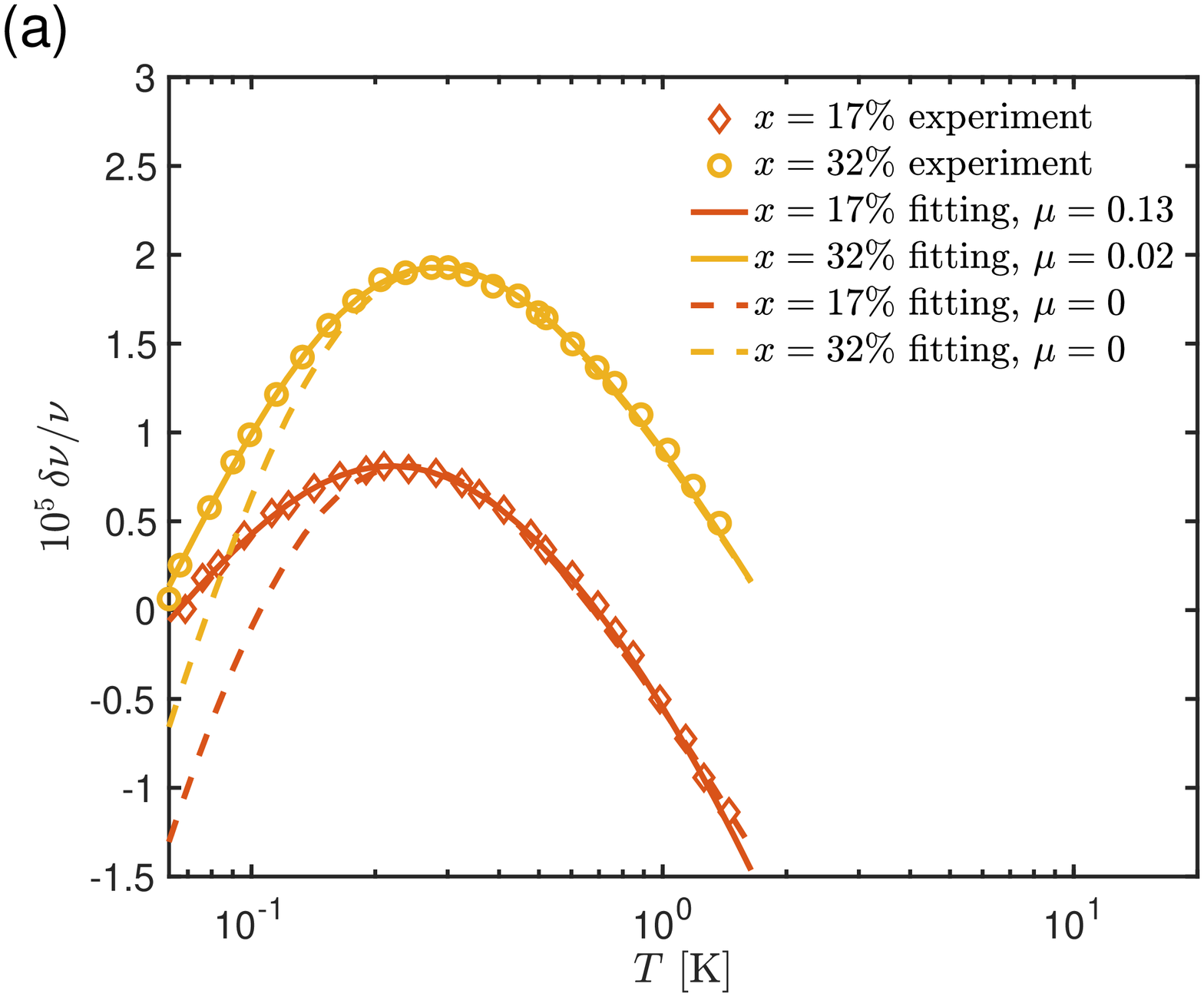}
\includegraphics[width=0.5\textwidth,height=0.27\textheight]{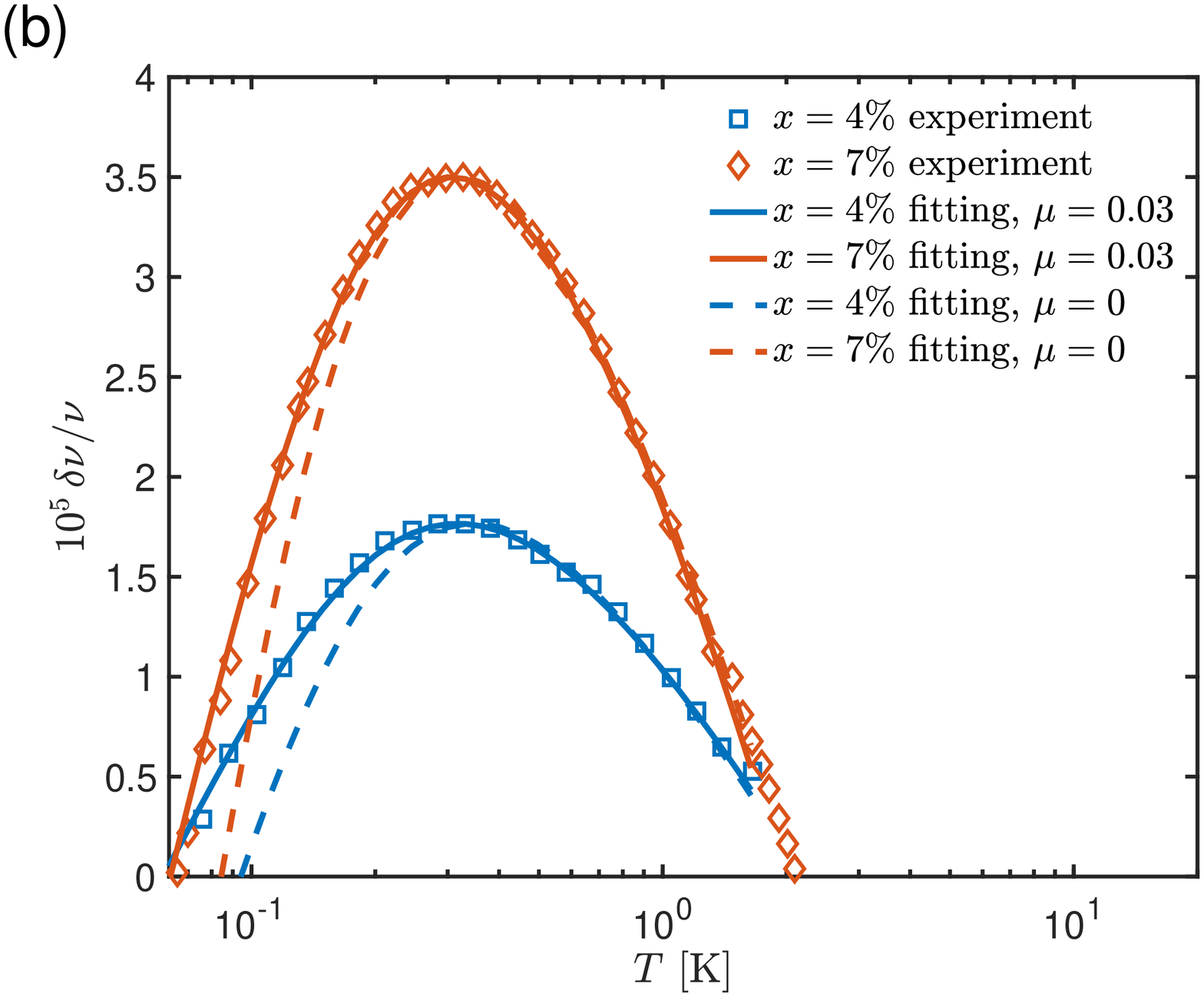}
\caption{(Color online) Temperature dependence of acoustic velocity.
Points denote experimentally obtained values, taken at $\omega=90$KHz \cite{TTP99}. Solid lines correspond to fits by the sum of Eqs.~(\ref{eq:res-gen}) and~(\ref{eq:relax-gen}), using Eq.~(\ref{approxDOS}) for the DOS with finite $\mu>0$ as a fitting parameter.
Dashed lines are best fits within the STM, i.e. with $\mu=0$. (a) For ${\rm (SrF_2)_{1-x}(LaF_3)_x}$.
Here as $x$ is enhanced so does the strain, and consequently the ratio of dipolar interaction to elastic interaction is decreased. Reduction of the power $\mu$ is in agreement with theory. Parameters used for best fits are
$P_0 \gamma^2/\rho v^2=1.9 \cdot 10^{-5}, 2.5 \cdot 10^{-5}$ and $A=7.9$MHz, $4.6$MHz, for $x=0.17, 0.32$ respectively.
(b) For ${\rm [(BaF_2)_{0.5}(SrF_2)_{0.5}]_{1-x}(LaF_3)_x}$. Here the strain, and thus the ratio of electric to elastic interactions, are independent of $x$. Independence of the power $\mu$ on $x$ is in agreement with theory. Parameters used for best fits are
$P_0 \gamma^2/\rho v^2=2.1 \cdot 10^{-5}, 4.5 \cdot 10^{-5}$ and $A=3.5$MHz, $3.9$MHz, for $x=0.04, 0.07$ respectively.
}
\label{fig:velocityTopp1}
\end{figure}

The form of the Hamiltonian (\ref{eq:Ising}) is equivalent to that of the STM Hamiltonian, i.e. the weakly interacting $\tau$-TLSs in the two-TLS model are equivalent to the TLSs in the STM. However, within the two-TLS model one can derive the typical magnitude of the interactions between the weakly interacting TLSs, as well as the typical magnitude of the random field. Since typical disorder energy at nearest neighbor distance is $\approx\gamma^{2}_{\mathrm{s}}/(\rho v^{2} R^{3}_{0})\sim T^{}_{\mathrm{g}}$, where $T^{}_{\mathrm{g}}\approx 300-1000\,$K is the glass transition temperature, one finds that the typical disorder energy for a $\tau$-TLS, which is $g$ times smaller, is given by $h^{}_0\approx 10\,$K~\cite{SM13,CA15,CBS13,CBS14}, and that TLS-TLS interactions at nearest neighbor distance have a typical value of $J^{}_{0}\approx gh^{}_{0}\approx 0.3\,$K $\ll h^{}_{0}$~\cite{SM13,CA15,CBS13}.

Consider now the electric dipolar interaction,
\begin{align}
\label{eq:Electric_interactions}&J^{}_{ij}\approx\frac{c^{}_{ij}p^{2}}{4\pi\epsilon\epsilon^{}_{0}R^{3}_{ij}}.
\end{align}
{\it The above characteristics of the weakly interacting $\tau$-TLSs within the two-TLS model, including the relative smallness of the random fields exerted on the $\tau$-TLSs, and the extreme smallness of the elastic $\tau$-TLS---$\tau$-TLS interaction, allow for the possibility of electric dipole interactions to dominate over the elastic $\tau$-TLS---$\tau$-TLS interactions, and to be not much smaller than the typical bias energies of the weakly interacting $\tau$-TLSs, i.e.\ $J^{}_{0}/h^{}_{0} \lesssim 1$.}

In Table~\ref{tbl} we summarize the typical energy scales of (a) interactions between general defects, which is of the order of the glass transition, (b) disorder energy and (c) typical TLS-TLS interaction energy - for the abundant ($\tau$-)TLSs at low temperatures. All energy scales are denoted within the two-TLS model with
(i) dominant elastic TLS-TLS interactions, and (ii) strong electric dipolar TLS-TLS interactions.

Experimentally, the ratio $r$ between the electric and elastic interactions (see App.~\ref{sec:AppB} for an exact definition) varies strongly between different amorphous materials. It can be deduced from combined measurements of dielectric loss and acoustic loss on the same material. Such measurements were carried out for vitreous silica, BK7 and coverglass~\cite{NRO98}, indicating a value of $r=0.3$ for vitreous silica and $r=1.51$ for BK7 and coverglass (see the detailed analysis in App.~\ref{sec:AppB}). While all amorphous materials display slope ratios of roughly $1:-1$ in the temperature dependence of the velocity and dielectric constant, the detailed functional behavior differs between materials, and corresponds to quite different energy dependence of the TLS-DOS, as displayed in Fig.~\ref{fig:approxDOS}. Indeed, fitting the data for the acoustic velocity in vitreous silica and in coverglass, we find a smaller power $\mu$ in the former in comparison to the latter.

For some other amorphous materials, the above ratio $r$ cannot be inferred directly from experiments. Yet, it can be approximated by measured values of the acoustic interaction constant and the dipole moment of TLSs in these materials. While for vitreous silica the measured dipole moment is small ($0.5\,$D~\cite{GSHD79}), typical values of $\gamma^{}_{\mathrm{w}}\approx 1-2\,$eV~\cite{GB76,HA94,GGJ12,PAP14} and the corresponding parameters (mass density, sound velocity and dielectric constant), and dipole moment values $p\approx 3-6\,$D~\cite{MJM05,PAP14,LJ15,DTC13,KMS14,SB16} of selected amorphous solids (including amorphous solids relevant to modern superconducting qubits and microresonators, such as Al$^{}_{2}$O$^{}_{3}$ and Si$^{}_{3}$N$^{}_{4}$ \cite{MCL19}), suggest that values $r$ larger than unity may be present in these materials. In turn, that would correspond to a power-law dependence of the TLS-DOS at low energies with $\mu \geq 0.1$.

The relation suggested above between the magnitude of the power of the energy dependence of the TLS-DOS and the dominance of the electric dipolar interaction over the elastic interaction can be further examined by considering disordered lattices, where the TLS concentration can be varied \cite{PLC99}. Two protocols exist for the variation of the concentration of TLSs \cite{Watson95,TTP99}. In the first, TLSs are the sole defects in the lattice. In this case, increasing the TLS concentration increases the strain in the system, and thus the coupling of $\tau$-TLSs to the phonon field \cite{SM08,SM13}. In the second, the TLS concentration is varied in a mixed lattice, in which strain is large already in the absence of TLSs. Thus, in the second protocol elastic and electric dipole interactions strengthen equally with an increased TLS concentration, as a result of the reduced typical distance between TLSs. However, in the first protocol, in addition to the reduced distance between TLSs, the increased strain with TLS concentration results in a decreased ratio of the electric dipolar to elastic TLS-TLS interactions. This is reflected in the temperature dependence of the sound velocity as fitted with the DOS of Eq.~(\ref{approxDOS}), with a smaller power $\mu$ as the TLS concentration is increased [Fig.~\ref{fig:velocityTopp1}(a)]. However, in the mixed crystal plotted in Fig.~\ref{fig:velocityTopp1}(b) strain is large and independent of TLS concentration, leading to a similar and small value of $\mu$ at TLS concentrations of $4\%$ and $7\%$.

\section{Summary}
\label{Sec:Summary}

We have considered TLSs in amorphous solids for which their mutual interactions are not much smaller than the randomness in their bias energies.
Such a scenario emerges naturally within the two-TLS model, provided that electric interactions dominate over elastic interactions. Data for BK7 and coverglass \cite{NRO98} attest for larger electric dipolar than elastic interactions in these materials, and typical parameters for amorphous solids used in superconducting resonators suggest that a stronger dominance of the electric dipolar interactions may be expected.
Our results clearly indicate the relation between the strength of the TLS-TLS interaction strength and the temperature dependence of the thermal conductivity, acoustic velocity, and dielectric constant in amorphous solids and disordered lattices.
While more comprehensive data and their analysis is desirable, our results seem to support the origin of such relatively strong TLS-TLS interactions lying in the dominance of electric over acoustic TLS-TLS interactions in certain materials.
We thus suggest a possible microscopic origin for the power-law dependence of the single-particle TLS-DOS at low energies, as found experimentally, and for the resulting anomalous exponents of the low-temperature thermal conductivity and specific heat and the temperature dependence of the acoustic velocity and dielectric constant at low-temperatures.

Energy-dependent TLS-DOS, albeit weaker, is obtained also within the dipolar gap theory of the STM \cite{Burin95}. A comprehensive study of the relation between the acoustic and dielectric responses in various amorphous solids could examine whether systems in which dipolar interactions dominate over elastic interactions are abundant, as well as the relevance of the dipolar gap theory and the two-TLS model in describing the low-energy properties of amorphous solids at low temperatures.

\begin{acknowledgments}
SM acknowledges support from the Alexander von Humboldt and Minerva foundations. AB and AM acknowledge partial support by the National Science Foundation (CHE-1462075). AB acknowledges the support of LINK Program of NSF and Louisiana Board of Regents and MPIPKS (Dresden, Germany) Visitor Program in 2018 and 2019. MS acknowledges support from the Israel Science Foundation (Grants No. 821/14 and No. 2300/19).
\end{acknowledgments}

\appendix

\section{Calculation of the thermal conductivity and of the acoustic velocity}
\label{sec:AppC}

Our numerical calculation of the TLS-DOS was performed under the approximation of a zero tunneling amplitude. Deviations resulting from a finite tunneling amplitude are expected to be significant only at the lowest energies, and to have only small, logarithmic corrections to the physical quantities of interest in this work, and thus no qualitative effect on our results.

\begin{figure}[ht!]
\includegraphics[width=0.5\textwidth,height=0.27\textheight]{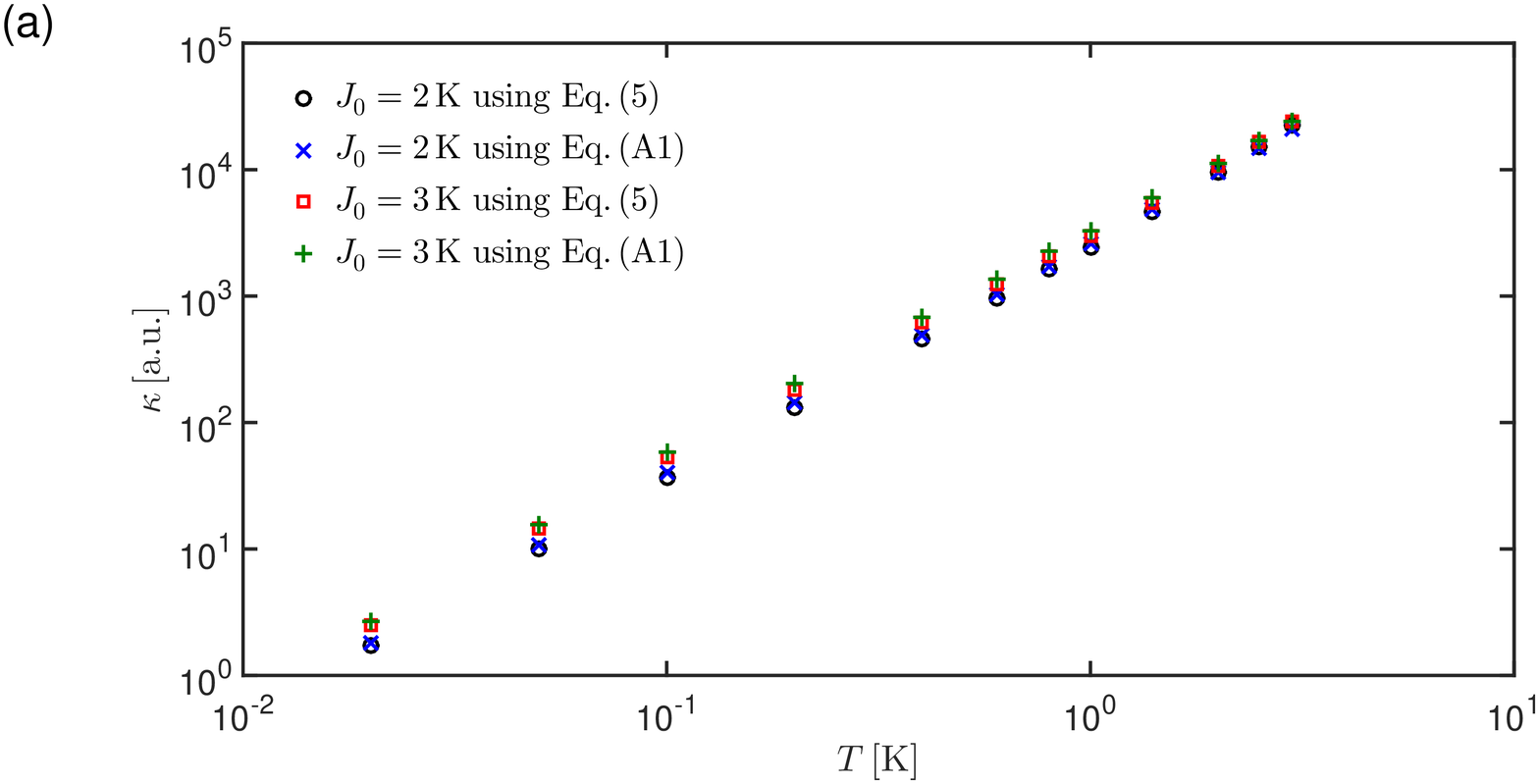}
\includegraphics[width=0.5\textwidth,height=0.27\textheight]{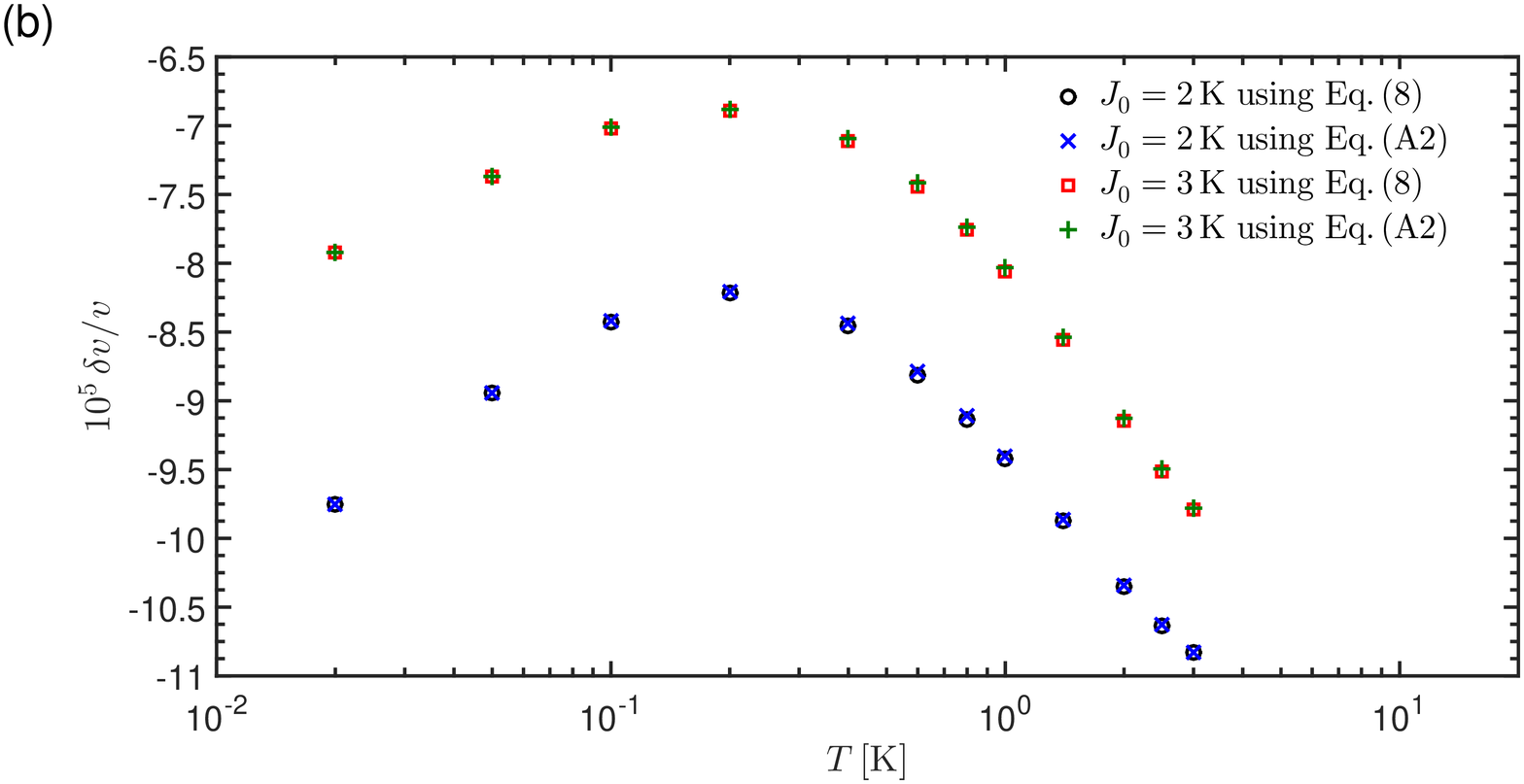}
\caption{(Color online) (a) Thermal conductivity calculated numerically using (i) Eq.~(\ref{eq:kappa1}) as in the main text, assuming the numerically calculated TLS DOS to be $n(E)$, and (ii) Eq.~(\ref{eq:kappa2}) derived in App.~\ref{sec:AppC}, assuming the numerically calculated TLS DOS to be $n(\Delta)$. Plots demonstrate negligible differences between the two approximations. (b) Similarly, negligible differences are found for the sound velocity calculated numerically using (i) Eq.~(\ref{eq:relax-gen}) as in the main text, and (ii) Eq.~(\ref{eq:relax-genDelta}) derived in App.~\ref{sec:AppC}.
}
\label{fig:numericalverification}
\end{figure}

With regard to the thermal conductivity, the calculation in the main text was performed under the assumption that $E=\Delta$. An alternative approach would be to perform the integral over $\Delta^{}_{0}$ in Eq.~(\ref{eq:5}), assuming that $P^{}_{0}(\Delta)$ is proportional to the numerically calculated TLS-DOS. The resulting expression for the thermal conductivity is:
	
\begin{align}
\label{eq:kappa2}\kappa(T)\propto\int^{\infty}_{0}\frac{E^{4}dE}{T^{2}\sinh^{2}(E/2T)\tanh(E/2T)\int^{E}_{0}P^{}_{0}(\Delta)d\Delta} \, .
\end{align}
We now replace $P^{}_{0}(\Delta)d\Delta$ by $n(T,E')dE'$. This equation thus transforms into Eq.~(\ref{eq:kappa1}) by replacing the integral in the denominator by its value at its upper limit. Since phonon energies of order temperature dominate the thermal conductivity, and since $n(T,E')\approx n(E=T)$ for $E'\leq T$, we expect thermal conductivities calculated via Eq.~(\ref{eq:kappa2}) and Eq.~(\ref{eq:kappa1}) to differ negligibly, as is indeed verified numerically (see Fig.~\ref{fig:numericalverification}).

Similar considerations apply to the calculation of the sound velocity and equivalently to the calculation of the dielectric constant. Considering the sound velocity, Eq.~\ref{eq:res-gen} for the resonant contribution remains intact. For the relaxation contribution one obtains

\begin{align}
\frac{\delta v^{}_{\mathrm{rel}}}{v}
=&-\frac{1}{L^{}_{0}}\frac{\gamma^2}{\rho v^2}\int_{0}^{\infty} \frac{dE}{2T\cosh^2\left(E/2T\right)}
\nonumber\\
&\times\int_{0}^{1} \frac{\sqrt{1-x^2}dx}{x}\frac{n(T,E\sqrt{1-x^2})}{1+\frac{\omega^2}{\left[Ax^2E^3\coth\left(E/2T\right)\right]^2}}.
\label{eq:relax-genDelta}
\end{align}
Note that by replacing $n(T,E\sqrt{1-x^2})$ with $n(T,E)$ one obtains again Eq.~(\ref{eq:relax-gen}) in the main text.
Indeed, we have repeated the calculation of the thermal conductivity and acoustic velocity.
Here too, using the numerically calculated TLS-DOS and Eqs.~(\ref{eq:res-gen}) and (\ref{eq:relax-genDelta}), we re-obtain the results in Fig.~\ref{fig:numericalvelocity} up to negligible differences.
We emphasize that neither interpretation of the numerically calculated TLS-DOS, i.e. as $n(E)$ (main text) or $n(\Delta)$ (here) is exact, since the latter assumes for $n(\Delta)$ a value obtained by considering the Efros-Shklovskii argument for the full energies.

\section{Two-TLS model}
\label{sec:AppA}

While Hamiltonian~(\ref{eq:Ising}) is standard in the theory of interacting TLSs~\cite{BJL77},
we now derive it as a low-energy effective Hamiltonian of the two-TLS model~\cite{SM13}. This allows the determination of the energy scales of the TLS-TLS interactions and of the random fields, and the ratio between the two energy scales for both cases of elastic dominated and electric dominated TLS-TLS interactions.

The two-TLS model was microscopically derived \cite{SM13} and thoroughly validated  ~\cite{SM08,GS11,CBS13,CBS14,GGS15} for disordered lattices which share the same universal low-temperature phenomena with amorphous solids~\cite{PRO02}. Here we assume the generalized validity of the two-TLS model in describing TLSs in amorphous solids. Such a viewpoint is supported by the following: (i) TLS-dictated low-energy properties show the same universal phenomena in disodered lattices and in amorphous solids~\cite{PRO02}, and careful experimental work suggests that universality in both groups of systems is of the same origin~\cite{YKMP86,LVP+98,PLC99} (ii) The two-TLS model derives the smallness and universality of phonon attenuation as is given by the universal dimensionless ``tunneling strength"~\cite{APW72,PWA72,PRO02} for both disordered lattices and amorphous solids (iii) The two-TLS model was found useful in explaining TLS pure dephasing and nonlinear absorption in superconducting qubit and microresonator circuits, not accounted for by the STM~\cite{LJ16,MS16,KSB+17}.

First considering only elastic interactions, the two-TLS model divides TLSs into two groups, with bimodal distribution of their interaction strengths with the strain, leading to the TLS-phonon interaction Hamiltonian~\cite{SM13}
\begin{align}
\label{eq:TLSphonon}&\mathcal{H}^{}_{\mathrm{int}}=-\sum^{}_{i}\sum^{}_{\alpha,\beta}\left(\eta^{}_{i}\delta^{}_{\alpha\beta}+\gamma^{i}_{\mathrm{w},\alpha\beta}\,\tau^{z}_{i}+\gamma^{i}_{\mathrm{s},\alpha\beta}\,S^{z}_{i}\right)\varepsilon^{i}_{\alpha\beta}.
\end{align}
Here, $\varepsilon^{i}_{\alpha\beta}$ is the strain tensor at TLS $i$, and the sum over $\alpha$ and $\beta$ runs over the three Cartesian coordinates $x$, $y$ and $z$. $\tau$-TLSs possess an intrinsically small interaction with the strain, and as a result weak TLS-TLS interactions~\cite{SM13,SM08,GS11,CBS13,CBS14,GGS15}. Such TLSs constitute the predominant degrees of freedom at low energies~\cite{SM13,CA15,CBS14}; their state is described by the Ising variable $\tau^{z}_{i}$ ($\tau^{z}_{i}=\pm 1$) and its weak coupling to the strain is given by the tensor $\gamma^{i}_{\mathrm{w},\alpha\beta}$. $S$-TLSs are described by the Ising variables $S^{z}_{i}$ ($S^{z}_{i}=\pm 1$) and are coupled much more strongly to the strain field, with $|\gamma^{i}_{\mathrm{s},\alpha\beta}|\sim\gamma^{}_{\mathrm{s}}$, where $g\equiv\gamma_{\mathrm{w}}/\gamma_{\mathrm{s}}\approx 0.02$~\cite{SM13,SM08,GS11,CBS13,CBS14,GGS15}. The first term of Eq.~(\ref{eq:TLSphonon}) describes a volume energy due to the strain field which is independent on the orientation of defects, where usually $\eta^{}_{i}\lesssim\gamma^{}_{\mathrm{s}}$.

The density of states (DOS) of $S$-TLSs strongly diminishes at low energies~\cite{SM13,CBS14,CA15}, and at low temperatures, for most purposes, these TLSs can be treated as frozen variables having no dynamics. They then contribute an additional term to the energy of the same order of magnitude
$\left(\propto\gamma^{}_{\mathrm{s}}\right)$ as the volume term. By integrating out the phonon amplitudes, at lowest order perturbation theory, one obtains the Hamiltonian~(\ref{eq:Ising}) as the low-energy effective Hamiltonian of the system~\cite{SM13}, with typical random fields and interactions as are given in Eq.~(\ref{eq:Random_field_and_interactions}) in the main text.

\section{Magnitude of the elastic and electric interactions}
\label{sec:AppB}

The parameters $c^{}_{i}$, $c^{}_{ij}\sim O(1)$ in Eq.~(\ref{eq:Random_field_and_interactions}) contain the angular dependence of the random field and the interaction.
Generally, the $c_{ij}$ parameters have a complicated dependence on the relative orientation and position of the TLSs, and can be regarded as normally distributed random variables~\cite{SM08}. Under this assumption one can establish a connection between the interaction strength expressed as the dimensionless parameter $P^{}_{0}\braket{|\tilde{J}^{}_{0,\mathrm{el}}|}$ for the elastic interaction and $P^{}_{0}\braket{|\tilde{J}^{}_{0,\mathrm{dip}}|}$ for the electric dipolar interaction, assuming that only terms with the transverse sound velocity $v^{}_{t}$ are significant for the elastic interaction. Here $\tilde{J}^{}_{0,\mathrm{el}}$ and $\tilde{J}^{}_{0,\mathrm{dip}}$ are interaction constants for elastic and electric interactions, respectively.

Then for the elastic interaction one can express the internal friction as
$Q^{-1} = \frac{\pi}{2}P^{}_{0}\gamma^{2}_{0}/(\rho v^{2}_{t})$, where $\gamma_0^2$ is the average squared of the off-diagonal component of the TLS-strain interaction constant tensor, and the logarithmic slope of the temperature dependence of the sound velocity in the resonant regime as $C^{}_{\mathrm{el}} = P^{}_{0}\gamma^{2}_{0}/(\rho v^{2}_{t})$~\cite{NRO98}. The average absolute value of TLS-TLS interaction constant\cite{BAL98,SM08} has been evaluated numerically assuming independent Gaussian distributions of elastic tensor components, similarly to Ref.~\onlinecite{abecho}, and it can be expressed as $P^{}_{0}\braket{|\tilde{J}^{}_{0,\mathrm{el}}|}\approx 1.1 Q^{-1}= 1.74C^{}_{\mathrm{el}}$. Similar analysis for the electric dipolar interaction yields $P^{}_{0}\braket{|\tilde{J}^{}_{0,\mathrm{dip}}|}\approx 0.36\tan\delta=0.56C^{}_{\mathrm{dip}}$, where $C^{}_{\mathrm{dip}}$ is the slope of the logarithmic temperature dependence of the dielectric constant. Consequently, the ratio of the two averaged interaction constants $r\equiv\braket{|\tilde{J}^{}_{0,\mathrm{dip}}|}/\braket{|\tilde{J}^{}_{0,\mathrm{el}}|}$ can be expressed as $r=0.36\tan\delta/(1.1 Q^{-1})=0.56C^{}_{\mathrm{dip}}/1.74C^{}_{\mathrm{el}}$. Using the available experimental data for loss tangent and internal friction we find for SiO$_2$ that elastic interactions dominate ($C^{}_{\mathrm{el}}=0.17-0.23\cdot 10^{-3}$\cite{SRTO94}, $Q = 3.3\cdot 10^{-4}$~\cite{CBEH00}, yields $r=0.3$); yet for BK7 and coverglass~\cite{NRO98} we find dominant electric dipolar interactions (BK7: $C^{}_{\mathrm{dip}}=1.51\cdot 10^{-3}$, $C^{}_{\mathrm{el}}=3.2\cdot 10^{-4}$ yields $r=1.51$, coverglass: $C^{}_{\mathrm{dip}}=1.69\cdot 10^{-3}$, $C^{}_{\mathrm{el}}=3.6\cdot 10^{-4}$ also yields $r=1.51$). It would be of much interest to further investigate experimentally the magnitudes of the elastic and electric TLS-TLS interactions in amorphous solids. We further note that our estimates above neglect variations at short range of the elastic interaction and of the dielectric constant, both of which may enhance the magnitude of the electric TLS-TLS interactions in comparison to the magnitude of the elastic TLS-TLS interactions and in comparison to the elastically dominated typical random fields.

\section{Fitting parameters for the approximate equation for the TLS-DOS}
\label{sec:AppD}

In Table II we detail the values of $B(T)$ and $\mu(T)$ obtained by fitting the function form in Eq.~(\ref{approxDOS}) to the numerical simulated data of $n(E,T)$. For each temperature fitting was performed by requiring best fits in the energy range $0<E<2T$.

\begin{table}[ht!]
  \begin{center}
    \label{tab:table2}
    \begin{tabular}{l|c|r}
      \multicolumn{3}{c}{$J^{}_{0}=2\,$K} \\
      \hline \hline
      \textbf{$T$} & \textbf{$B(T)$} & \textbf{$\mu(T)$}\\
      \hline
      0.1 & 0.0118 & 0.103\\
      0.2 & 0.0126 & 0.119\\
      0.4 & 0.0133 & 0.122\\
      0.5 & 0.0135 & 0.127\\
      0.6 & 0.0137 & 0.126\\
      0.8 & 0.0140 & 0.104\\
      1    & 0.0142 & 0.118\\
    \end{tabular}
\qquad
    \begin{tabular}{l|c|r}
      \multicolumn{3}{c}{$J^{}_{0}=3\,$K} \\
      \hline \hline
      \textbf{$T$} & \textbf{$B(T)$} & \textbf{$\mu(T)$}\\
      \hline
      0.1 & 0.0077 & 0.081\\
      0.2 & 0.0090 & 0.141\\
      0.4 & 0.0097 & 0.151\\
      0.5 & 0.0100 & 0.152\\
      0.6 & 0.0102 & 0.152\\
      0.8 & 0.0105 & 0.165\\
      1    & 0.0108 & 0.158\\
    \end{tabular}
\caption{$B(T)$ and $\mu(T)$ values for $J^{}_{0}=2,3\,$K.}
  \end{center}
\end{table}

\end{document}